\documentclass[sigconf]{acmart}
\AtBeginDocument{%
  }

\setcopyright{cc}
\setcctype{by}
\copyrightyear{2026}
\acmYear{2026}
\acmDOI{10.1145/3795867.3831010}

\acmConference[SIGCSE Virtual 2026]{Proceedings of the 2nd ACM Virtual Global Computing Education Conference V.1}{November 12--15, 2026}{Virtual Event, USA}

\acmBooktitle{Proceedings of the 2nd ACM Virtual Global Computing Education Conference V.1 (SIGCSE Virtual 2026), November 12--15, 2026, Virtual Event, USA}

\acmISBN{979-8-4007-2506-7/2026/11}
\usepackage{tabularx}
\usepackage{booktabs}

\begin{document}

\title{LearnAI: Just-in-Time AI Co-Creation Across Disciplines at a University}


\author{Weihao Qu}
\orcid{0000-0003-1027-6556}
\affiliation{%
  \institution{Monmouth University}
  \city{West Long Branch}
  \state{NJ}
  \country{United States}
}
\email{wqu@monmouth.edu}

\author{Ling Zheng}
\orcid{0009-0005-1702-9787}
\affiliation{%
  \institution{Monmouth University}
  \city{West Long Branch}
  \state{NJ}
  \country{United States}
}
\email{lzheng@monmouth.edu}

\author{Chris Buzaid}
\orcid{0009-0000-6034-6917}
\affiliation{%
  \institution{Monmouth University}
  \city{West Long Branch}
  \state{NJ}
  \country{United States}
}
\email{s1363246@monmouth.edu}

\author{Daniel Crawford}
\orcid{0009-0004-0649-5355}
\affiliation{%
  \institution{Monmouth University}
  \city{West Long Branch}
  \state{NJ}
  \country{United States}
}
\email{s1323702@monmouth.edu}


\renewcommand{\shortauthors}{Qu et al.}
\newcommand{\THESYSTEM}{LearnAI}

\begin{abstract}
As generative AI reshapes professional and educational practice, institutions face a challenge: how to support diverse learners, from non-coders to advanced students, in building confidence and practice with AI-supported problem solving. Most institutional responses bifurcate into conceptual workshops for general audiences or technical courses for computer science majors, leaving few spaces where mixed-ability learners can engage common AI tasks at levels matched to their prior experience.

This experience report presents the LearnAI Framework, a two-layer model for just-in-time AI co-creation piloted at a comprehensive teaching university. The Wide-Exposure Layer embeds short presentations in existing courses to build AI awareness at scale, reaching students and faculty across 18 courses in five disciplines. The Customized Co-Creation Layer provides opt-in, one-on-one sessions where clients work with trained undergraduate tutors through a 5-Stage Pedagogical Script: Problem Framing, Tool-Task Mapping, Iterative Co-Prompting, Deployment and Verification, and Ethical Reflection.
Over two semesters, 35 clients co-created 36 portfolio websites and over 20 deployed web applications.
Interviews with five clients and two tutors suggest a recurring change in how clients described AI use, shifting from treating AI as a passive answer machine to engaging it as a collaborative tool under human direction. A small paired pre/post AI readiness dataset ($N=7$) provides preliminary descriptive context, and tutor accounts document how the pedagogical script was enacted and adapted across client types.
We report on boundary cases including clients who felt overwhelmed and respondents who deliberately rejected AI use. This paper contributes a practical, adoptable framework with initial evidence from a single institution.
\end{abstract}





\maketitle

\section{Introduction}
\label{sec:intro}
Generative AI is rapidly reshaping how people across all backgrounds approach problem-solving~\cite{Yao_2025,Shulgina_2025}. However, institutional responses in higher education remain slow. Universities have moved quickly on policies and licenses, but pedagogy lags behind~\cite{Jin_2025}. 
This lag matters because students and faculty already confront immediate, situated problems that require computational thinking yet often fall outside traditional CS curricula, such as building portfolio websites or automating workflows. Without guided, just-in-time support, these authentic needs remain missed opportunities to teach core computing competencies, including requirement decomposition, algorithmic verification, and system deployment, as educational moments rather than mere IT requests~\cite{Robe_2022,Edouard_2022}.

\begin{figure}[h]
    \centering
    \includegraphics[width=0.98\linewidth]{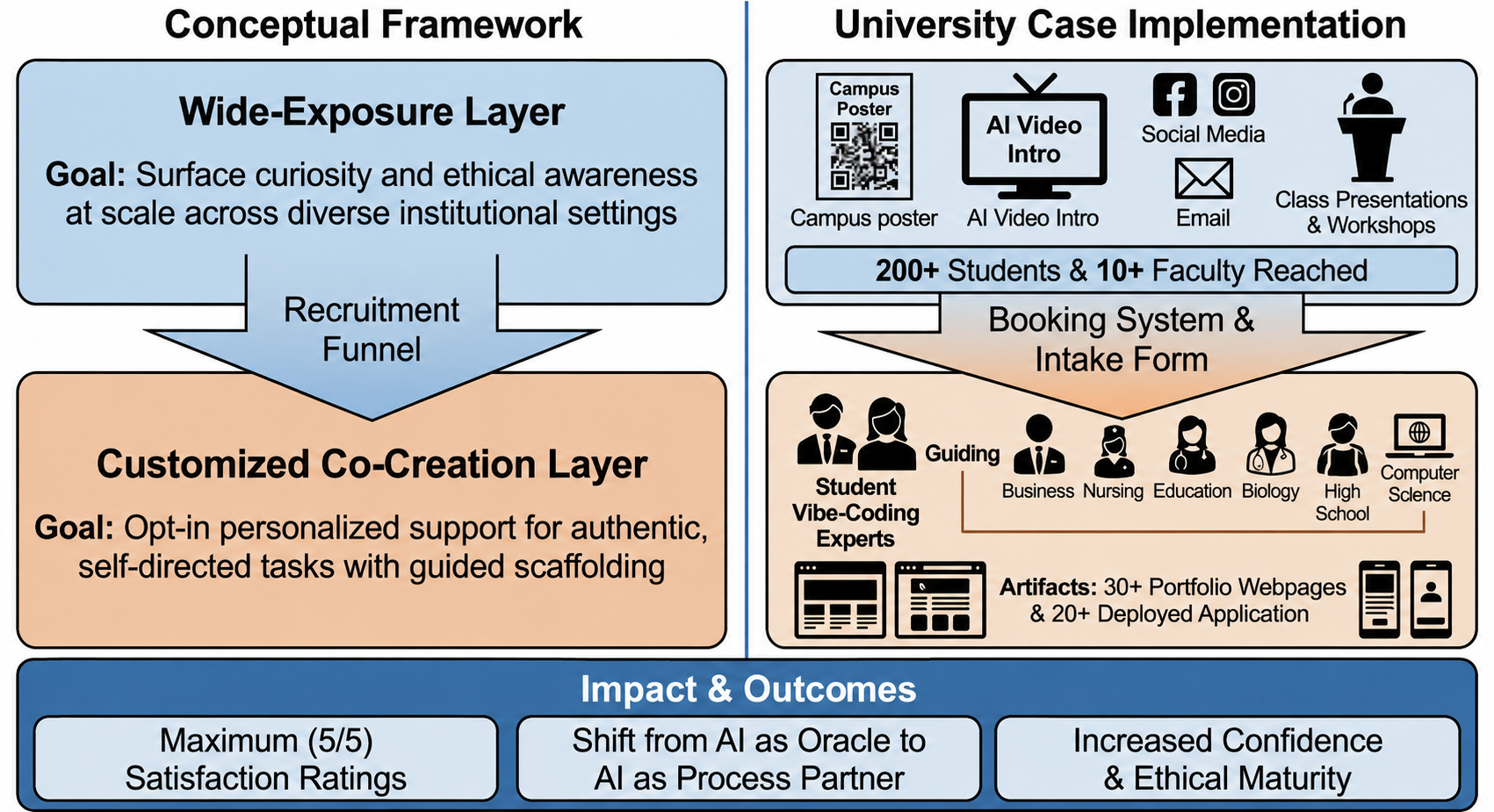}
    \caption{The {\THESYSTEM}  Framework. The left column outlines the general design pattern, while the right column details the implementation at a comprehensive teaching university.}
    \Description{A two-column framework diagram showing LearnAI's wide-exposure layer connected by recruitment and intake to a customized co-creation layer, with implementation examples and outcomes.}
    \label{fig:overall}
\end{figure}

This experience report presents the {\THESYSTEM} Framework, a two-layer model for just-in-time AI co-creation piloted at a comprehensive university.
The framework addresses a practical question: how can a single institution support mixed-ability learners, from individuals with no coding experience to advanced computer science majors, in building \emph{AI readiness}: confidence, awareness of appropriate use, and ability to apply AI tools within authentic workflows?
It does so through two complementary layers: a Wide-Exposure Layer that builds awareness at scale, and a Customized Co-Creation Layer that provides intensive, 1-on-1 scaffolding for authentic tasks in  Figure~\ref{fig:overall}. Rather than teaching programming syntax, {\THESYSTEM} emphasizes problem framing, tool orchestration, critical verification of outputs, and situated ethical decision-making~\cite{Romero_2025,Nally_2025}.

We report on an instantiation during Fall 2025 through Spring 2026 with students, faculty, and staff whose backgrounds ranged from no coding experience to advanced CS majors.
The Wide-Exposure layer reached 18 courses (293 enrolled students), and the Co-Creation layer enrolled 35 clients who co-created 36 portfolio websites and over 20 deployed applications. Our analysis combines Layer~1 survey responses, artifact inspection, a small paired readiness dataset ($N=7$), and interviews with five clients and two tutors. The paired readiness scores are reported as preliminary descriptive context. Interviews, tutor accounts, artifacts, and boundary cases provide the primary basis for tracing a shift from treating AI as a passive answer machine (``Oracle'') toward engaging it as a human-directed collaborative tool (``Process Partner'').
At the same time, several respondents rejected AI use entirely; we frame this as an intentional educational philosophy rather than a knowledge gap.


\section{Related Work}
\label{sec:relatedwork}

\paragraph{AI Literacy and Mixed-Ability Learning}
AI literacy encompasses competencies for critically evaluating AI, using AI tools responsibly, and understanding societal implications~\cite{long2020ai,Zhang_2025}. Most initiatives, however, bifurcate into conceptual workshops for general audiences or technically demanding courses for CS majors~\cite{Biagini_2024,becker2023generative}, creating few spaces where beginners and advanced learners practice shared competencies at appropriate depths. In computing education, recent work has begun integrating AI literacy into curricula through programming exercises~\cite{Horvath_2022,Aldemir_2025,Galmar_2024,Curi_2025} or critical analysis of AI systems~\cite{Saltz_2019}, but these efforts typically target a single population. In mixed-ability institutions, AI readiness therefore must extend beyond conceptual awareness or programming-course content to include situated problem-solving across roles, disciplines, and prior experience. This motivates frameworks that operationalize AI literacy as problem-solving capability in mixed-ability settings~\cite{Ng_2021,Salma_Fatih_Tofiq_2025}, treating ethical concerns and questions about authorship and intellectual ownership~\cite{Reza_2025,Draxler_2024} as central pedagogical content rather than policy constraints~\cite{Gaidartzi_2025}.

\paragraph{Peer Tutoring and Studio-Based Computing Education}
LearnAI also builds on peer tutoring and studio-based models in computing education. Peer tutoring shows how structured, role-based interaction can support learner-to-learner help~\cite{Topping_1996}, while studio-based computing education emphasizes authentic project work, critique, collaboration, and design artifacts~\cite{Hundhausen_2008}. LearnAI adapts these traditions for AI co-creation by having undergraduate tutors use a shared script with student, faculty, and staff clients working on self-selected tasks.

\paragraph{Just-in-Time AI Authorship and Institutional Infrastructure}
Research on AI-assisted programming shows that generative AI enhances productivity for experienced practitioners while lowering entry barriers for novices~\cite{terragni2025future,sarsa2022automatic}, but learners must cultivate new competencies---code evaluation, diagnostic debugging, and architectural decision-making---that span human and AI-generated components~\cite{Pang_2025,perry2023users}. A Just-in-Time (JiT) approach~\cite{Stavredes_2005,Yilmaz_2022} delivers instruction at moments of authentic need~\cite{Contel_2024} and can position non-specialists as active authors engaged in genuine system design via sustained human facilitation~\cite{Romero_2025}. This emphasis on facilitation aligns with human-AI interaction guidelines that foreground user understanding, correction, and control~\cite{amershi2019guidelines}, framing the Oracle-to-Process-Partner shift as movement from accepting AI output toward directing, checking, and revising it. Beyond learner-level design, adoption of generative AI at the institutional level has outpaced curricular integration~\cite{Jin_2025,Xiao_2023}. Within this institutional gap, the {\THESYSTEM} Framework operates as a service-oriented infrastructure complementing traditional curricula~\cite{Ram_2025,Paidic_n_Soto_2024}, positioning ethics as lived pedagogical content negotiated through authentic task engagement~\cite{Xu_2025,Alsharefeen_2025}.

\section{The LearnAI Framework \& Implementation}
\label{sec:method}
This study was conducted at a comprehensive, teaching-focused university in North America during Fall 2025 through Spring 2026. The intervention targeted a mixed-ability population ranging from undergraduates to tenured faculty and administrative staff. The study drew on two overlapping pools: students across 18 courses (total enrollment 293) reached through broad outreach and 15 faculty from various departments, and a subset of 35 clients who opted into intensive co-creation. The study protocol was approved by the university's Institutional Review Board (IRB), with informed consent obtained for all data collection.



\subsection{Layer 1: Wide-Exposure and Recruitment}

The wide exposure layer embedded short, context-specific presentations provided by an undergraduate AI expert team (4 tutors) in 18 existing courses across introductory IT, science, business, and nursing, supported by asynchronous outreach through social media and campus publicity. Fifteen-minute class presentations demonstrated practical AI use patterns aligned with immediate student needs: study support, concept explanation, research assistance, and portfolio site creation, followed by discussion of ethical constraints, hallucinations, citation practices, and institutional AI policies.

Selected sessions concluded with an online survey capturing consent, affiliation, self-reported AI use, prompting confidence, perceived learning impact, and barriers to use. Survey respondents could opt in for follow-up resources and one-on-one sessions, enabling the wide-exposure layer to function as both instruction and intake.

\subsection{Layer 2: Customized Co-Creation Labs}


The customized co-creation layer comprised opt-in one-hour sessions scheduled through a dedicated booking system, in which clients worked with a tutor on a self-selected task using AI tools. Clients enrolled through an online intake form that captured role, department, self-reported AI experience, and a brief task description such as building a portfolio site, automating workflows, or developing study tools or web applications.

\begin{figure}
    \centering
    \includegraphics[width=0.98\linewidth]{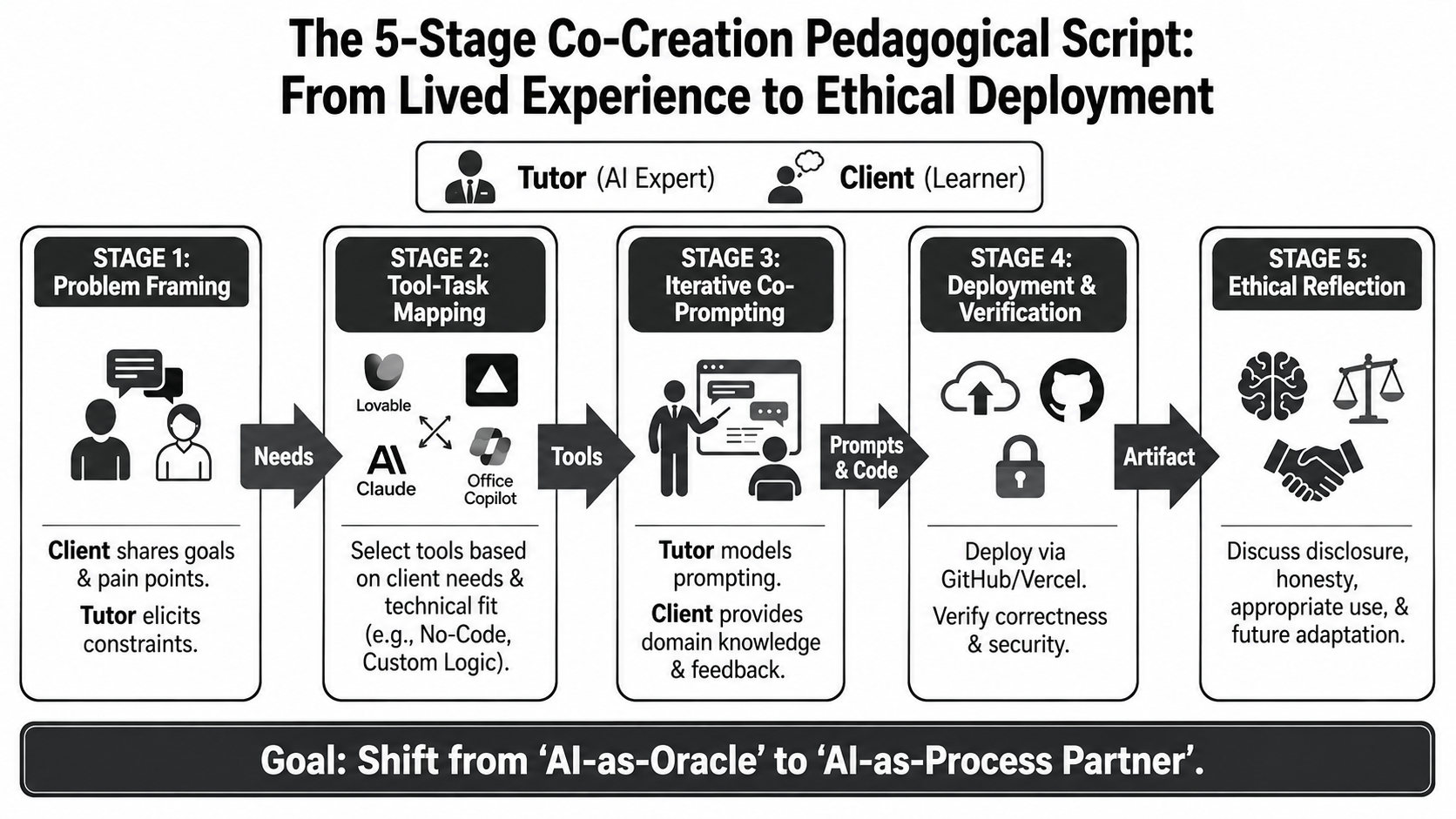}
    \Description{Flow diagram showing clients moving through problem framing, tool-task mapping, co-prompting, deployment and verification, and ethical reflection.}
    \caption{Just-in-Time Co-Creation Process Flow in Customized Co-Creation Layer.
    }
    \label{fig:cocreation}
\end{figure}

\paragraph{The 5-Stage Pedagogical Script.}
Sessions followed a structured protocol (Figure~\ref{fig:cocreation}) designed to shift clients from an ``Oracle'' mindset (expecting answers) to a ``Process Partner'' mindset (co-creating solutions). The script provides a repeatable structure adapted to the client's background and task:

\textbf{Stage 1: Problem Framing} (5--10 min). Tutors ask ``What problem are you trying to solve?'' rather than ``What do you want AI to do?'' to elicit concrete requirements: audience, constraints, and success criteria. Tutors withhold tool suggestions until the problem is clearly scoped. A common pitfall was that clients arrived with vague goals, such as wanting ``a website,'' which required additional time for scoping.
\textbf{Stage 2: Tool-Task Mapping} (3--5 min). The tutor and client select tools based on task fit and constraints rather than default familiarity~\cite{wang2024farsight,Kim_2024}. Tutors had to avoid defaulting to a single familiar tool for every client, since tool choice depended on background, access constraints, and task type.
\textbf{Stage 3: Iterative Co-Prompting} (20--30 min). Tutors model ``prompting as specification'': writing prompts that read like requirements documents. The tutor gradually releases control, moving from demonstrating to coaching to observing. A recurring challenge was helping clients move past the expectation that AI would ``just do it'' and instead treat prompting as an iterative specification process.
\textbf{Stage 4: Deployment \& Verification} (10--15 min). Sessions move beyond the chat window to actual deployment (e.g., Vercel, GitHub Pages). Tutors teach clients to verify outputs against requirements and check for hallucinated features, guarding against accepting plausible AI output without testing whether promised features actually worked.
\textbf{Stage 5: Ethical Reflection} (3--5 min). After testing the artifact, tutors helped clients decide how it should be used, shared, or limited in practice. This discussion connected the client's task to privacy, hallucination checking, course or workplace rules, attribution, and human ownership of final decisions. Because debugging or deployment sometimes consumed the final minutes, this stage was the easiest to compress. We therefore reframed it in facilitator guidance as a required session-end checklist.

After each session, tutors produced a facilitator-verified summary documenting tools used and next steps and shared with clients to support independent continuation. Clients completed a pre-session and post-session AI readiness assessment (described below) and were invited to provide open-ended feedback.

\subsection{Facilitator Preparation}
Layer~2 sessions were facilitated by a team of four undergraduate tutors, all computer science majors. Tutors were selected based on demonstrated proficiency with generative AI tools (Claude, Copilot, Vercel) and experience building deployed applications. Before facilitating sessions, tutors completed a calibration process: they observed two sessions led by the faculty coordinator, then co-facilitated two sessions before leading independently. The four tutors collectively facilitated over 70 sessions across two semesters. The 5-Stage Script provided a shared structure, but tutors adapted pacing and tool selection to individual clients. Semi-structured interviews with two of the four tutors: one senior tutor (30+ sessions facilitated) and one junior tutor ($\sim$5 sessions), documented how the script was enacted in practice and what adaptations they made for non-technical versus technical clients (see Section~\ref{sec:finding}). The two layers formed a reinforcement loop: wide-exposure sessions recruited clients into co-creation, while completed artifacts and ethical negotiations fed back into later presentations as local exemplars.

\subsection{Data Collection and Analysis}
We employed a mixed-methods design triangulating quantitative data with qualitative accounts. Data sources included:
\begin{description}
    \item[Survey Data (Layer 1):] Layer~1 included 18 courses with a combined enrollment of 293. Because survey administration was instructor opt-in, the survey was administered in 7 of those courses (estimated combined enrollment of approximately 120 students) spanning CS, software engineering, business, IT, and nursing. We received 58 responses, an estimated 48\% participation rate among students in survey-receiving courses; 54 respondents consented to research use. The survey captured usage frequency, prompting proficiency, and attitudes toward AI.
    \item[Pre/Post AI Readiness Assessment (Layer 2):] We used six scenario-based items covering AI Foundations, Prompt Engineering, Critical Evaluation, Ethics \& Safety, Human-AI Collaboration, and Tool Selection. Each item was scored from 1 to 4, summed to 24 points, and converted to a percentage. The post-assessment included a 5-point Likert confidence item. Seven clients provided usable paired responses. The instrument is preliminary and has not yet undergone formal validation, so we treat AI readiness as an operational construct rather than a validated psychometric measure.
    \item[Semi-Structured Interviews:] Interviews with five Layer~2 clients (purposively sampled for diversity in role, session count, and coding background; all at least four months post-first-session) and two of the four tutors.
    \item[Artifact Corpus:] The artifact corpus included 36 portfolio websites and over 20 deployed web applications. Inspection recorded whether artifacts had an active link or demonstration, matched the stated task, and included task-relevant deployment features such as authentication, persistence, administrative controls, or public hosting. This inspection was informal and was not a validated quality rubric. Selected public examples appear on the official LearnAI site (\url{https://www.lainow.com/}) when permission, consent, and privacy review allow. Private or operational tools are described by function rather than linked.
\end{description}

Because this was an exploratory deployment, we used the survey and readiness scores to describe what happened rather than to prove impact. The paired comparison reported in Section~\ref{sec:finding} is included for context only. The five client interviews and two tutor interviews were reviewed repeatedly to identify recurring ideas. These ideas were grouped around the Oracle/Process Partner shift, stage-level engagement, and themes that emerged from the interviews. We read the artifacts, self-reports, and interviews together to understand patterns across the deployment, while recognizing that the study was opt-in and conducted at a single institution. We present the findings by moving from Layer~1 reach and attitudes to Layer~2 readiness and interviews, then to artifacts and boundary cases.


\section{Findings}
\label{sec:finding}
\subsection{Quantitative Reach and Attitudes}

Post-session survey data ($N=54$ consented respondents from 7 of 18 courses) from Layer~1 indicated that while most respondents utilized AI tools daily (35.2\%) or weekly (48.1\%), usage was largely passive (e.g., summarization). Respondents self-rated as ``intermediate prompters'' and valued specific workflows like quiz generation, yet open-ended responses revealed persistent ethical anxiety. Frequent descriptions of AI as ``cheating'' or fears of hallucination underscored the gap between passive consumption and confident authorship, consistent with the need for the deeper scaffolding provided in Layer 2.

\subsection{Pre-Post AI Readiness Assessment (Layer 2)}
In the Customized Co-Creation Layer, the lab supported 35 unique clients. Seven clients provided usable paired pre-post assessment responses. Scores increased from pre-session ($M=82.5\%$, $SD=15.5\%$) to post-session ($M=89.9\%$, $SD=17.5\%$), a mean gain of 7.4 percentage points. Five of seven clients improved, one remained at ceiling, and one declined. Two clients advanced one readiness level (L2: Developing User→L3: Proficient Creator and L3→L4: Process Partner).
Because the paired sample is small ($N=7$) and the readiness instrument is preliminary, we treat these results as descriptive rather than inferential. We report the pre/post scores to document observed responses only, not to claim a measured intervention effect. Post-session self-reported confidence ratings were high descriptively ($M=4.7/5$, all ratings $\geq 4$). Artifact production (36 portfolios, 20+ deployed apps) reflects task completion, not measured learning gains.

\subsection{Qualitative Evidence of Cognitive Shift}
Semi-structured interviews with five Layer~2 clients and two tutors suggested a common theme: interviewees described a shift in how they understood AI's role. Before sessions, several clients framed AI in passive, oracle-like terms. After sessions, they described AI as a collaborative tool requiring human direction.
\paragraph{Client Interview}
A CS professor who built five applications across her sessions initially described AI use as limited to ``ChatGPT, ask questions and do some travel plans.'' After sessions, her framing shifted:
\begin{quote}
\small\itshape
``It can ask me questions that, to think about the angles that I didn't think about it\ldots So from that point, I will feel that it's more like a collaborator.''
\end{quote}

A professor in software engineering initially described AI as ``something [for] chatty questions.'' After six sessions building a lab tracker:
\begin{quote}
\small\itshape
``I would say maybe a collaborator. Somebody, something to help me do something.''
\end{quote}

An administrative staff member who used AI only ``in a chat kind of way\ldots enhanced Google'' reported that after two sessions she now looks ``to automate more of the administrative tasks'' and teaches colleagues to ``end every prompt with `ask me questions to complete the task,'\,''  as discussed in Section~\ref{christy}.

An education student with no coding background captured the affective dimension:
\begin{quote}
\small\itshape
``I can't imagine AI can help me do such a lot of things\ldots I can build up my personal website within 10 minutes or even five minutes.''
\end{quote}
She now combines Claude, VS Code, and GitHub to create presentation slides, a workflow she independently applied to a conference talk that impressed colleagues, as discussed in Section~\ref{bee}.
\paragraph{Tutor Interview}
Tutor interviews were consistent with this pattern. The senior tutor (30+ sessions) described a structured approach:
\begin{quote}
\small\itshape
``I have a script that I follow\ldots the first part is I try to figure out exactly what problem we're trying to solve today. Also I'm trying to figure out their current AI knowledge.''
\end{quote}
He emphasized that good tutoring means fostering independence:
``A good tutor is someone that teaches the student to go off on their own as soon as possible and not need to book another session ever again: it's like the anti-business model.''

 The junior tutor ($\sim$5 sessions) described adapting tool selection to client comfort:
 \begin{quote}
\small\itshape
 ``If they're more comfortable without any code, I wouldn't recommend Claude Code\ldots it might be better to go for something like a no-code tool like Lovable.''
 \end{quote}
 Both interviewed tutors identified communication as the hardest skill to develop in new tutors, above technical proficiency.
Notably, more than 10 clients returned for multiple sessions or initiated secondary independent projects, suggesting continued engagement with AI-supported workflows, though longer-term follow-up is needed to assess durability.

While these outcomes are encouraging, not all experiences were uniformly positive. At least one client without a technical background reported feeling overwhelmed by the number of tools and steps involved in building a website (e.g., GitHub, hosting, multiple platforms) and expressed a need for more foundational scaffolding, highlighting a boundary condition for the current design. Survey data also document a subset of students (3) who reject AI in their education or feel that the overhead of prompting outweighs potential benefits, underscoring that the framework primarily serves those who opt in and are at least tentatively open to AI.
Artifacts then complement interviews by showing what clients produced and deployed. Table~\ref{tab:evidence} summarizes how each major claim is supported by the available data sources and where limitations apply.

\begin{table}[h]
\scriptsize
\centering
\caption{Evidence Map: Claims, Data Sources, and Limitations}
\label{tab:evidence}
\begin{tabularx}{\columnwidth}{@{} X X X @{}}
\toprule
\textbf{Claim} & \textbf{Supporting Evidence} & \textbf{Limitation} \\ \midrule
Readiness score change & Pre/post assessment (7 paired responses: 5 improved, 1 remained at ceiling, 1 declined) & Preliminary; underpowered; descriptive only. \\ \addlinespace
Oracle$\to$Process Partner shift & Client interviews (5), tutor interviews (2) & Self-report, purposive sample, retrospective recall \\ \addlinespace
Deployed artifacts & Artifact corpus (36 portfolios, 20+ apps inspected for access ) & Informal inspection; no quality rubric. \\ \addlinespace
Mixed-ability clients can co-create & Vignettes across 6 disciplines, Table~\ref{tab:tool-mapping} & Self-selected clients; no nonparticipant comparison \\ \addlinespace
Ethical anxiety in Layer~1 & Survey ($N=54$) & Instructor opt-in; self-report \\ \addlinespace
Clients develop AI boundaries & Client interviews (writing, privacy) & 1--2 interviewees; may not generalize \\ \addlinespace
Replication guidance & Tutor interviews (2): communication $>$ technical skill & Only 2 of 4 tutors; single institution \\ \bottomrule
\end{tabularx}
\end{table}

\subsection{Artifacts and Deployment}
Clients co-created a diverse array of functional software artifacts, summarized in Table~\ref{tab:tool-mapping}. These ranged from domain-specific tools (e.g., attendance trackers, peer grading platforms) to creative applications (e.g., cybersecurity games~\cite{Li_2025}, portfolio sites) and workflow automation (e.g., Excel data cleaning on locked-down hardware).
 Table~\ref{tab:tool-mapping} illustrates the range of artifact types and inspected deployment features; it should not be read as scores from a formal artifact-quality rubric.

\begin{table}[h]
\scriptsize
\centering
\caption{Summary of Tool-Task Mapping by Discipline and Project Scope}
\label{tab:tool-mapping}
\begin{tabularx}{\columnwidth}{@{} p{1cm} X X X @{}}
\toprule
\textbf{Discipline} & \textbf{Authentic Task / Pain Point} & \textbf{Toolset (Mapping)} & \textbf{Deployed Outcome} \\ \midrule
Nursing Faculty & Departmental newsletter with faculty login. & \textbf{Lovable} (No-code builder). & Live platform using institutional email. \\ \addlinespace
CS Faculty & Manual attendance management across 5 courses. & \textbf{Claude}, \textbf{Google AI Studio}, \textbf{Vercel}. & Secure QR-based system used by 50+ students. \\ \addlinespace
Business Student & Startup prototype requiring user data storage. & \textbf{Claude}, \textbf{Supabase}. & Functioning web app with persistent backend. \\ \addlinespace
Education (Grad) & Professional portfolio \& email refinement tool. & \textbf{Claude}, \textbf{GitHub}, \textbf{Vercel}. & Live portfolio and self-directed web app. \\ \addlinespace
Admin Staff & Repetitive data processing. & \textbf{Office Copilot}, \textbf{Excel} (Integrated AI). & Automated data cleaning and reporting workflows. \\ \addlinespace
SE Faculty & Redesigning robotics course using vibe coding. & \textbf{Claude}, \textbf{Gemini API}, \textbf{Python}. & Robotics curriculum with AI dialogue \\ \addlinespace
CS Student & Web-based game for cybersecurity project. & \textbf{Claude}, \textbf{Vercel}. & Interactive learning game used for classwork. \\ \addlinespace
CS Faculty & Peer grading platform for group presentation . & \textbf{Claude, Vercel, Supabase}. & Web application for class use. \\ \bottomrule
\end{tabularx}
\end{table}

\subsection{Representative Vignettes}

\subsubsection{Domain-Specific Tools}
\paragraph{Nursing Faculty: Departmental Newsletter as First AI-Mediated Tool}
 A nursing faculty member with no coding experience sought to build a departmental newsletter. Initially frustrated that a no-code builder (Lovable) failed to "understand" her needs, she worked with tutors to reframe her prompts as precise technical specifications. By the second session, having learned to decompose the system logic and verify the AI's output, she successfully deployed a live application with secure faculty login. This case demonstrates how the co-creation model shifts focus from syntax to requirement specification, allowing domain experts to author their own infrastructure.


\paragraph{Computer Science Faculty: Attendance Tracker and Beyond}
A computer science faculty member sought to replace a manual attendance workflow but lacked the time to develop a custom solution from scratch. In two sessions, he used Claude to scaffold a web application and Vercel for immediate deployment. The co-creation process focused on high-level architecture: implementing secure university-authenticated login and dynamic QR codes to ensure integrity. The system was adopted by three faculty and used in five courses with over 50 students in Fall 2025. Most importantly, this experience "unblocked" his development process; he subsequently independently built and deployed a course-specific learning game, demonstrating how the co-creation model accelerates prototyping even for technical experts. This case shows how the framework enables transfer once AI supported workflows for design and deployment are established.

\subsubsection{Creative Applications}
\label{bee}
\paragraph{Education Graduate Student: From Portfolio to Email-Polisher App}
An education master's student with no coding background initially reported, ``I felt so overwhelmed when they talked too much about the code things,'' especially when sessions involved GitHub and VS Code, which she was ``still figuring out how to use.'' Tutors reduced this friction by introducing Lovable as a no-code entry point while gradually helping her connect Claude, VS Code, GitHub, and Vercel into a working development workflow. Across more than ten sessions, she used tutor-provided summaries to revisit each session and consolidate what the tools were doing. By the follow-up interview, she had independently recombined these tools to build materials for a conference talk presentation. She also recommended LearnAI to a peer who asked how she had created the presentation, suggesting movement from overwhelmed novice to persistent learner and independent builder.

\subsubsection{Workflow Automation }
\label{christy}

\paragraph{Administrative Staff: Workflow Automation on Locked-Down Hardware}
A departmental secretary on a locked-down desktop learned ``prompting as programming'' using Office Copilot across two sessions. In a follow-up interview four months later, she described independently chaining NotebookLM, Gamma, and HeyGen to automate department meeting notes: a multi-step workflow she developed on her own. She also reported that her son, an advertising major initially skeptical of AI, attended one session and subsequently built his own website and apps using GitHub, Supabase, and VS Code. This case shows adaptation under IT constraints and spillover beyond the direct client.

\section{Discussion}
\label{sec:discussion}

\subsection{From "Oracle" to "Process Partner"}

\begin{figure}[h]
    \centering
    \includegraphics[width=0.86\linewidth]{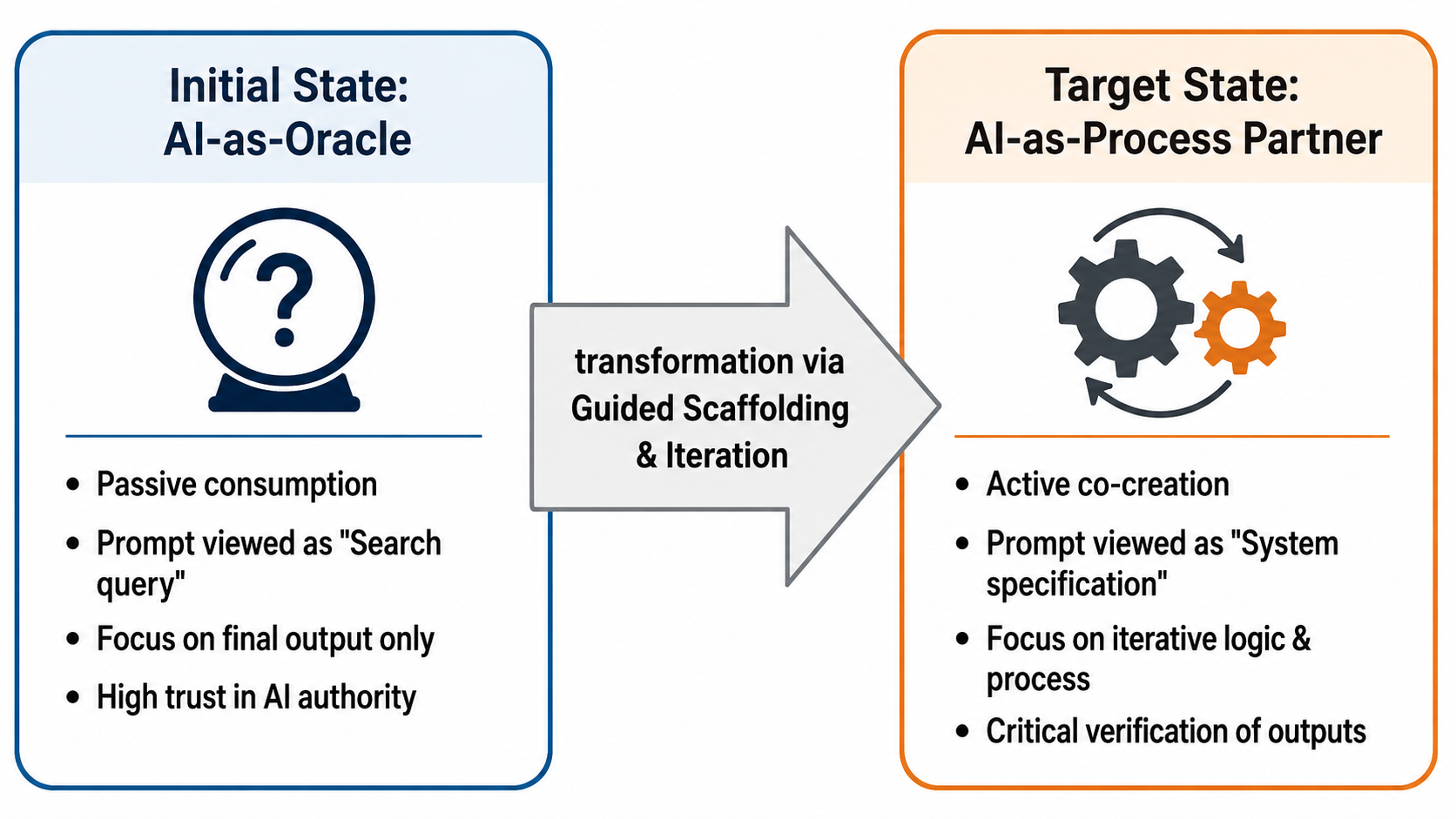}
    \Description{Diagram contrasting AI as an Oracle answer engine with AI as a Process Partner for planning, implementation, verification, and ownership.}
    \caption{Cognitive Shift in Learner Identity.}
    \label{fig:edu}
\end{figure}

A key observation across sessions is a change in how clients describe AI. As Figure~\ref{fig:edu} illustrates, many clients initially approached AI as an ``Oracle’’ or answer engine, engaging in passive consumption (e.g., ``Write this code for me’’). The co-creation script appeared to support a transition to a ``Process Partner’’ mindset: successful clients began to treat AI as a junior developer that could generate syntax while they retained ownership of architecture, requirements, and verification. This suggests that AI co-creation shifts cognitive load from syntax recall to requirements engineering and verification. The small pre/post readiness dataset provides descriptive context only; the shift claim rests primarily on interview accounts, tutor accounts, artifacts, and boundary cases.
\subsection{Mechanisms of Success}

Several mechanisms may help explain the framework's outcomes. Human facilitation amplified AI use by helping clients compose, debug, verify, and manage cognitive load in real time. Task-driven sessions sustained investment by beginning with each client's concrete problem, while follow-through pathways converted brief exposure into intensive co-creation. Concerns about cheating and hallucinations became task-specific instructional content rather than session-level barriers. Finally, matching tools to client constraints, rather than enforcing a fixed workflow, helped reduce resistance to AI use and work around institutional IT restrictions.

\subsection{Reframing Authorship and Boundaries}
Clients emphasized making artifacts themselves rather than delegating to AI. The framework treated prompt design, task decomposition, and verification as substantive intellectual labor, reframing what counts as ``doing the work’’ in AI-supported settings. At the same time, clients did not simply become more enthusiastic adopters. Some retained preferences for non-AI approaches or identified clear boundaries. A data science graduate student described an authorship boundary around scholarly writing:
\begin{quote}
\small\itshape
``I prefer to write papers by myself\ldots if you’re using AI from the beginning, it did give you a beautiful structure, but sometimes I may feel lose myself.’’
\end{quote}
He also identified privacy as a practical constraint, noting, ``If you’re handling some sensitive data, you probably don’t want AI to touch those information.’’ These boundaries complicate a purely adoption-oriented view of AI literacy: successful learning included knowing when AI could expand ideas and accelerate implementation, and when personal voice, disciplinary judgment, or data sensitivity should limit its role.

\subsection{Transferability of the Co-Creation Findings}
Several constraints bound these claims. Because Layer~2 was opt-in through course presentations, outreach, and booking links, the 35 clients should not be read as a conversion rate from the 293 students reached in Layer~1. Technical failures, multi-tool overload, incomplete Stage~5 logs, limited demographic data, and seven paired readiness responses make the results descriptive rather than established effects. The findings transfer most directly to learners with concrete needs and tentative openness to AI, and to institutions with peer tutors, faculty coordination, and lightweight deployment support.

\subsection{Implications for Research and Practice}
A two-layer, service-oriented lab can complement course-based AI education by offering just-in-time, project-centered support using existing peer tutors, widely available AI tools, and lightweight cloud platforms. For computing education research, the work illustrates how a small, high-touch lab can function as a site for studying AI-era problem-solving across mixed-ability populations, while foregrounding boundary cases such as non-adopters and overwhelmed clients as valuable evidence about the limits of productive complexity.
For institutions asking whether this model can be replicated, the two tutor interviews suggest that feasibility depends less on rare technical expertise than on tutor selection, pedagogy, and quality control. Both tutors identified communication and teaching as harder to develop than technical proficiency. Replication efforts should thus prioritize recruiting tutors who can explain and adapt, then add training on current tools and a readiness mechanism before independent facilitation.

\section{Conclusion}
\label{sec:conclusion}
The central lesson from LearnAI is that AI education can be treated as institutional practice rather than only as course content: broad exposure surfaces needs, while opt-in co-creation helps clients turn those needs into working artifacts. Across 18 courses and 35 clients, LearnAI produced 36 portfolio websites and over 20 deployed applications, while the 5-Stage Pedagogical Script documents how tutors helped clients move from vague goals to working artifacts. Because the paired readiness data are exploratory ($N=7$, preliminary instrument, no control group), the strongest evidence lies in artifact inspection, tutor accounts, client interviews, transfer cases, and boundary cases. Together, these cases suggest that co-creation can change how clients relate to AI while showing that adoption should remain negotiated rather than assumed. Future work should track clients longitudinally and release open-source materials including the assessment instrument, facilitator training guide, and session templates. Carefully replicated, this model points toward AI education embedded in institutional life, responsive to individual needs, and honest about its limits.

\clearpage
\bibliographystyle{ACM-Reference-Format}
\balance
\bibliography{main}










\end{document}